\documentclass[12pt,preprint]{aastex}

\newcommand{\al}{$A_\lambda$}
\newcommand{\ajk}{$A_J/A_K$}
\newcommand{\alk}{$A_{[\lambda]}/A_K$}
\newcommand{\ejk}{$E_{J-K}$}
\newcommand{\elk}{$E_{[\lambda]-K}$}
\newcommand{\ejhk}{$E_{J-H}/E_{H-K}$}
\newcommand{\gl}{GLIMPSE}

\newcommand{\lsim}{$l=$}
\newcommand{\mic}{$\mu$m}

\def\lesssim{\mathrel{\hbox{\rlap{\hbox{\lower4pt\hbox{$\sim$}}}\hbox{$<$}}}}
\def\gtrsim{\mathrel{\hbox{\rlap{\hbox{\lower4pt\hbox{$\sim$}}}\hbox{$>$}}}}

\begin{document}
\title{The Wavelength Dependence of Interstellar Extinction from
1.25 to 8.0$\mu$m Using GLIMPSE Data}
\author{ Indebetouw, R. \altaffilmark{1,2},
Mathis, J. S. \altaffilmark{1},
Babler, B. L. \altaffilmark{1},
Meade, M. R. \altaffilmark{1},Watson, C. \altaffilmark{1},
Whitney, B. A. \altaffilmark{3},
Wolff, M. J. \altaffilmark{3}, Wolfire, M. G. \altaffilmark{4},
Cohen, M. \altaffilmark{5},
Bania, T. M. \altaffilmark{6},
Benjamin, R. A. \altaffilmark{7}, Clemens, D. P. \altaffilmark{6},
Dickey, J. M. \altaffilmark{8}, Jackson, J. M. \altaffilmark{6},
Kobulnicky, H. A. \altaffilmark{9}, Marston, A. P. \altaffilmark{10},
Mercer, E. P. \altaffilmark{6}, 
Stauffer, J. R. \altaffilmark{11},
Stolovy, S. R. \altaffilmark{11}, Churchwell, E. \altaffilmark{1}}

\altaffiltext{1}{University of Wisconsin-Madison, Dept. of Astronomy,
475 N. Charter St., Madison, WI 53706\vspace{-0.6em}}
\altaffiltext{2}{current address: Dept. of Astronomy,
 P.O. Box 3818, University of Virginia,
Charlottesville, VA 22903\vspace{-0.6em}}
\altaffiltext{3}{University of Colorado, Space Science Institute,
1540 30th St., Suite 23, Boulder, CO 80303-1012\vspace{-0.6em}}
\altaffiltext{4}{University of Maryland, Astronomy Dept., College Park,
MD 20742-2421\vspace{-0.6em}}
\altaffiltext{5}{University of California-Berkeley, Radio Astronomy Lab,
601 Campbell Hall, Berkeley, CA 94720\vspace{-0.6em}}
\altaffiltext{6}{Boston University, Institute for Astrophysical
Research, 725 Commonwealth Ave., Boston, MA 02215\vspace{-0.6em}}
\altaffiltext{7}{University of Wisconsin-Whitewater, Physics Dept.,
800 W. Main St., Whitewater, WI 53190\vspace{-0.6em}}
\altaffiltext{8}{University of Minnesota, Dept. Astronomy, 116 Church
St., SE, Minneapolis, MN 55455\vspace{-0.6em}}
\altaffiltext{9}{University of Wyoming, Dept. Physics \& Astronomy,
PO Box 3905, Laramie, WY 82072\vspace{-0.6em}}
\altaffiltext{10}{ESTEC/SCI-SA,Postbus 299,2200 AG Noordwijk,The
Netherlands\vspace{-0.6em}}
\altaffiltext{11}{Caltech, Spitzer Science Center, MS 314-6, Pasadena,
CA 91125}

\begin{abstract}
\vspace{-1cm}
We determine and tabulate \alk, the wavelength dependence of
interstellar extinction, in the Galactic plane for
1.25\mic~$\le\lambda\le$~8.0\mic{} along two lines of sight:
\lsim~42\arcdeg{} and \lsim~284\arcdeg. The first is a relatively
quiescent and unremarkable region; the second contains the giant
\ion{H}{2} region RCW~49 as well as a ``field'' region unrelated to
the cluster and nebulosity.  Areas near these Galactic longitudes were
imaged at $J$, $H$, and $K$ bands by 2MASS and at 3--8\mic{} by {\it
Spitzer} for the \gl{} Legacy program.  We measure the mean values of
the color excess ratios $(A_{[\lambda]}-A_K)/(A_J-A_K)$ directly from
the color distributions of observed stars.  The extinction ratio
between two of the filters, e.g. \ajk, is required to calculate \alk{}
from those measured ratios.  We use the apparent $JHK$ magnitudes of
giant stars along our two sightlines, and fit the reddening as a
function of magnitude (distance) to determine $A_J$/kpc, $A_K$/kpc,
and \ajk.  Our values of \alk{} show a flattening across the
3--8\mic{} wavelength range, roughly consistent with the \citet{lutz}
extinction measurements derived for the sightline toward the Galactic
center.
\end{abstract}


%=========================================================================
\section{Introduction}
\label{intro}

Extinction by interstellar dust affects most astronomical
observations.  The wavelength dependence of interstellar extinction
has been studied extensively, but is not well-understood in the
3--9$\mu$m region \citep[][]{drainearaa}.  The InfraRed Array Camera
\citep[IRAC;][]{fazio04} on board the {\it Spitzer Space Telescope} is
in the process of vastly increasing the number of observations in this
wavelength region.  Understanding the effects of dust extinction in
the IRAC bands is important to properly interpret these observations.

The wavelength dependence of interstellar extinction, \al, is commonly
treated as ``universal'' in the infrared because it apparently varies
far less between different sightlines than does extinction in the
optical and ultraviolet.  Many authors have concluded that \al{} is a
power-law (\al~$\propto\lambda^{-\beta}$) between $\sim$1\mic{} and
$\sim$4\mic. \citet{mw90} found $\beta$=1.8 in the diffuse
interstellar medium (ISM) as well as the outer regions of the $\rho$
Oph and Tr\ 14/16 clouds.  Other authors have fitted values between
1.6 and 1.8 \citep{drainearaa}.  A value of $\beta$=1.8 implies
\ejhk=$(A_J-A_H)/(A_H-A_K)$=1.8$\pm$0.1 (the value changes slightly
depending on the exact filters and source spectrum).  Deep surveys of
specific dark clouds in the $JHK$ bands have revealed significantly
different color ratios, ranging from \ejhk=1.47$\pm$0.06 for luminous
southern stars \citep{he95} to 2.08$\pm$0.03 in the Coalsack
\citep[][and refs. therein]{racca}.  At wavelengths approaching
9.7\mic, \al{} is dominated by absorption by the Si-O stretching mode
of interstellar silicates.  There is uncertainty in the wavelength
dependence of interstellar extinction between the power-law regime at
1--2\mic{} and the discrete silicate feature. Observations of H$_{\rm
2}$ rovibrational lines in Orion \citep{bertoldi99,rosenthal00} are
consistent with continuation of the power-law to $\lesssim 4$\mic, but
their uncertainties are too large to constrain the extinction at
longer wavelengths.  {\it Infrared Space Observatory} (ISO)
observations of hydrogen recombination lines by \citet{lutz} towards
the Galactic center show a flattening of \al{} in the region
3\mic$\lesssim\lambda\lesssim$9\mic.  \citet{lutz99} confirms this
flattening using more recombination lines towards the Galactic center,
and also shows evidence of extra mid-IR extinction towards external
galaxies.

The Galactic Legacy Infrared Mid-Plane Survey Extraordinaire
\citep[\gl; see][]{pasp}, a {\it Spitzer} Legacy program, uses IRAC to
obtain photometric images of the inner Galactic plane in four filters
([3.6], [4.5], [5.8], and [8.0]\mic) simultaneously.  By combining
stellar photometry of the \gl{} images with the 2MASS\footnote{This
publication makes use of data products from the Two Micron All Sky
Survey (2MASS), which is a joint project of the University of
Massachusetts and the Infrared Processing and Analysis
Center/California Institute of Technology, funded by the National
Aeronautics and Space Administration and the National Science
Foundation. Throughout this paper we will refer to the 2MASS $K_s$
filter as $K$.}
point source catalog, we can sample {$A_{\lambda}/A_K$} at seven
wavelength points $[\lambda]$ in the near-infrared (NIR, defined in
this paper as 1.2--8.0\mic): 1.240, 1.664, 2.164, 3.545, 4.442, 5.675,
and 7.760\mic{}.  The adopted wavelengths are the isophotal
wavelengths of the 2MASS and IRAC filters convolved with a K2III
star, which we use as our standard probe of extinction along the line
of sight (see discussion below)\footnote{Filter wavelengths depend on
the spectrum of the source, but differences are small and have little
effect on our results - for example the flat-spectrum isophotal
wavelengths for 2MASS are 1.235, 1.662, and 2.159\mic{}, and the
SSC-provided isophotal wavelengths for IRAC are 3.535, 4.502, 5.650,
and 7.735\mic{}.}.

We describe the data and their basic reduction in \S\ref{data}, and
our method of measuring color excess ratios \elk/\ejk{} in
\S\ref{method}.  Section~\ref{discussion} discusses the conversion of
these colors to \al, and issues related to interpretation.  In
\S\ref{conclusions} we summarize and tabulate our recommended relative
extinction values for the GLIMPSE/IRAC bands.

%=========================================================================
\section{Photometry of \gl{} data}
\label{data}

\begin{figure}
\plotone{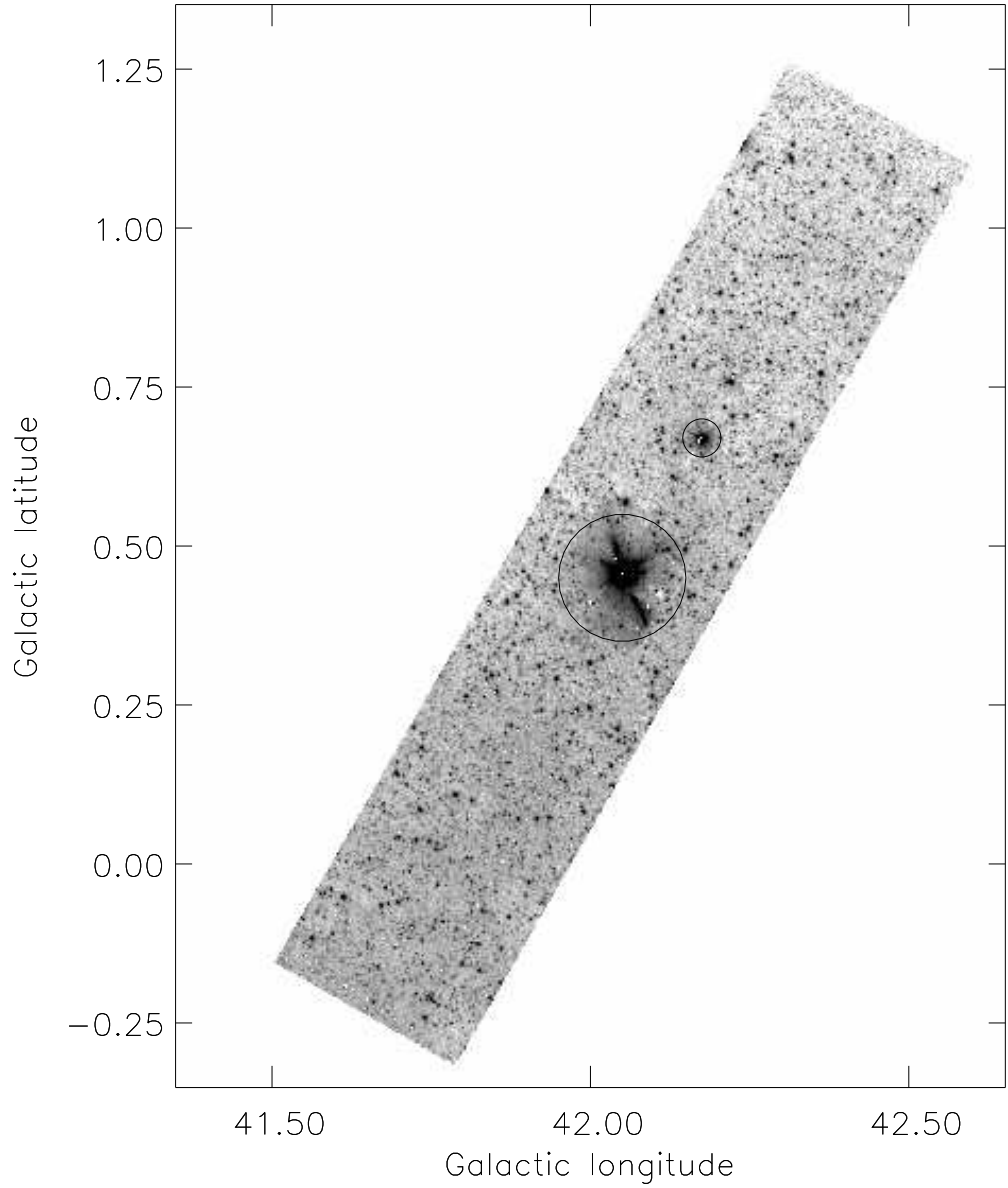}
\caption{\label{ioc} GLIMPSE 3.6\mic{} image of the region near
\lsim~42\arcdeg{} used in this study.  The bright spots in the center
are highly saturated stars; the regions around them are excluded from
\gl{} photometry (marked circles).}
\end{figure}

\begin{figure}
\epsscale{0.5}
\plotone{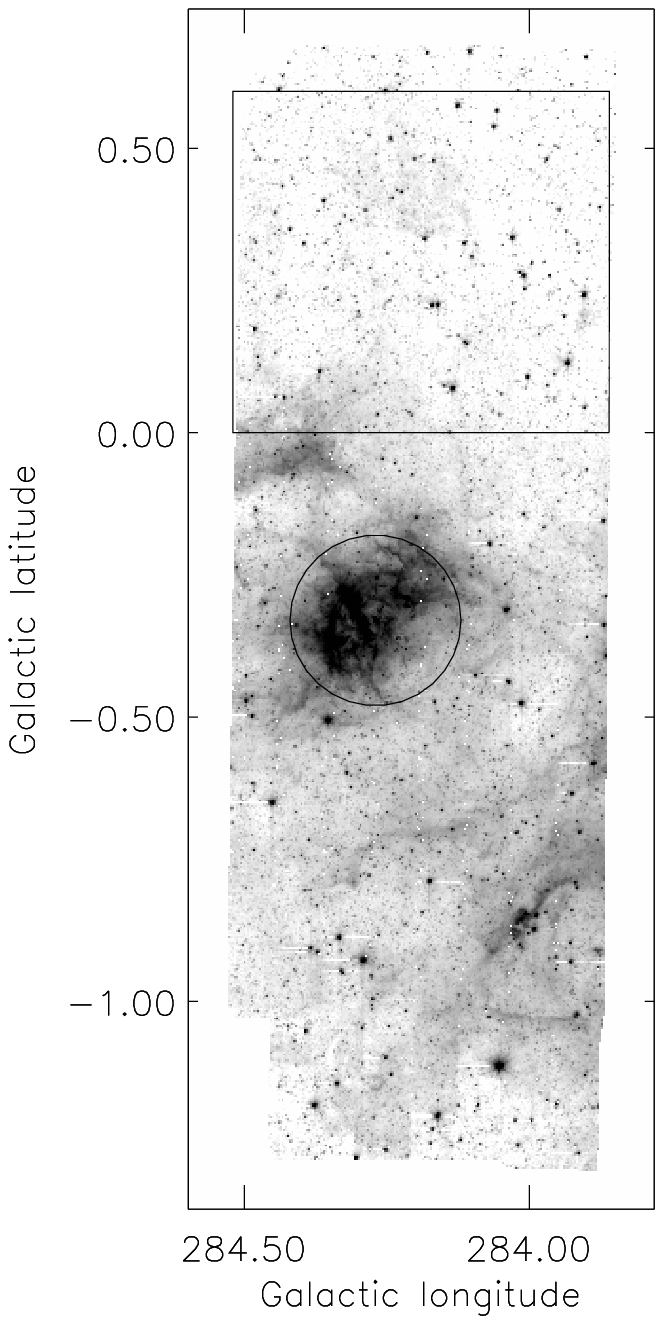}
\epsscale{1.0}
\caption{\label{osv} GLIMPSE 3.6\mic{} image of the region near
\lsim~284\arcdeg{} used in this study.  The region near the giant
\ion{H}{2} region RCW~49 is marked with a circle, and the ``field''
region away from nebulosity, star formation, and significant CO
emission \citep{dame} is marked with a square. }
\end{figure}

This paper uses stellar photometry in the Galactic plane obtained from
two subsets of \gl{} data.  The first dataset (Figure~\ref{ioc}),
0.\arcdeg33$\times$1.\arcdeg72 centered on ($l,b$)=(42\arcdeg,
0.5\arcdeg), was obtained during In-Orbit Checkout (IOC; October 2003,
PID 631).  The imaged region is a fairly unremarkable part of the
Galactic plane.  The second dataset (Figure~\ref{osv}),
283.9\arcdeg~$\le l \le$~284.5\arcdeg{} and -1.3\arcdeg~$\le b
\le$~0.7\arcdeg, was obtained for the purposes of Observing Strategy
Validation (OSV; December 2003, PID 195).  The region contains the giant
\ion{H}{2} region RCW~49 and its associated massive cluster
\citep[][and references therein]{ebc04}.

% FIGURE 1,2

Basic reduction of IRAC images was performed by the {\it Spitzer}
Science Center (SSC) Pipeline (ver. 8.9.0 at \lsim~42\arcdeg{} and
ver. 9.0.1 at \lsim~284\arcdeg).  Positional accuracies are better
than 1\arcsec{} \citep{werner}.  Point source full-width-half-max
resolution of IRAC data ranges from $\simeq$1.6\arcsec\ at [3.6] to
$\simeq$1.9\arcsec\ at [8.0].  The data were further processed by the
GLIMPSE pipeline \citep{pasp}.  Point sources were found and fluxes
extracted from each frame using a version of DAOPHOT \citep{stetson},
modified to be more robust in regions of rapidly varying high
background (Babler et al., in preparation).  To calculate stellar
colors, we adopt zero magnitude flux densities of 277.5, 179.5, 116.6,
and 63.1~Jy for the four IRAC bands (M.~Cohen 2004, private
communication).  For this present study, we culled the catalog to
include only those sources with signal-to-noise greater than 10 in
each band.  The magnitude to which stars are well-measured in all
bands is limited by the [5.8] and [8.0] bands that typically contain
high diffuse backgrounds (the calculated uncertainty for flux density
is higher if the local background is higher).  At the limiting
magnitude of 12.5, photometric uncertainties are less than 0.08, 0.09,
0.12, and 0.14 mag in the [3.6], [4.5], [5.8], and [8.0] bands,
respectively.  If we only require detections at wavelengths $<$5\mic{}
(e.g. for measuring the extinction at [4.5]), then we can include
sources as faint as 14~mag at [3.6] and [4.5], with photometric
uncertainties $<$0.15~mag.  The flux calibration was checked by
comparing extracted fluxes with modeled fluxes for five early A-type
dwarf stars -- these agreed to within 7\% for all IRAC bands
\citep{cohen03,kurucz93}.  Detections of the same source in different
images and different filters were bandmerged (cross-identified), and
detections were additionally merged with the 2MASS point source
catalog to provide $JHK$ fluxes for many sources.  Tests of
bandmerging in crowded fields using synthetic data show that fewer
than 0.5\% of sources are falsely cross-identified down to the flux
limits used in this work.

The data segment at \lsim~42\arcdeg{} was imaged before the telescope
was fully focused.  The point source function (PSF) is thus different
from that in subsequent IRAC data. We do not expect the precision of
our photometry to suffer since in the \gl{} pipeline the PSF is
constructed from the data for each epoch of data collection.  We did
not assess the flux calibration independently for the IOC data as we
did for the OSV data, but we do not see any gross errors, nor do we
expect any, because the standard {\it Spitzer} calibration strategy
was determined before launch.  Results in this paper are derived from
relative colors, which are insensitive to zero-point issues in the
photometry.

%=========================================================================
\section{Measurement of Extinction}
\label{method}

We determine the relative extinction \alk{} using photometry of
approximately 10,000 stars ($[\lambda]$ refers to the three 2MASS and
four IRAC filters).  Two different processes are required.  First, we
measure color excess ratios \elk/\ejk{} by fitting the loci of a
population of stars in color-color diagrams.  Secondly, we need a
single ratio of extinctions, say \ajk, determined or estimated
independently.  The calculation of \alk{} from \elk/\ejk{} follows
simply:
\begin{equation}
{{A_{[\lambda]}}\over{A_K}} =
{\left({{A_J}\over{A_K}}-1\right)}{{E_{[\lambda]-K}}\over{E_{J-K}}}+1\label{alamk}
\end{equation}
Various observational studies and dust theories find a range of
$A_J/A_K$ between 2.25 and 2.75 \citep{ccm89,fitz99,drainearaa}.  We
determine the ratio \ajk{} directly from our data by fitting the locus
of red clump giants in a color-magnitude diagram.  Both steps in our
process are only possible using a fairly sensitive, uniform, large
area survey such as 2MASS or GLIMPSE.

\subsection{Determining \ajk}
\label{jk}

We determine the extinction ratio \ajk{} by fitting the locus of
reddened K giants in color-magnitude space.  These stars are bright
enough \citep[$M_J \lesssim -1.5$, e.g.][]{lopez02} that 2MASS and
GLIMPSE can observe them to great distances through the Galactic
plane.  The intrinsic luminosity and color distributions of red clump
K giants is determined by the evolution of stars from the main
sequence and the initial mass function, is fairly narrow ($\pm
\lesssim 0.5$ mag) and should not change with distance.  We use the
wavelength-independent effect of distance on apparent magnitude (the
1/d$^2$ change in flux), together with the wavelength-dependent effect
of extinction, to determine the average extinction per unit distance,
and thus the absolute ratio of extinctions, \ajk.

\begin{figure}
\plotone{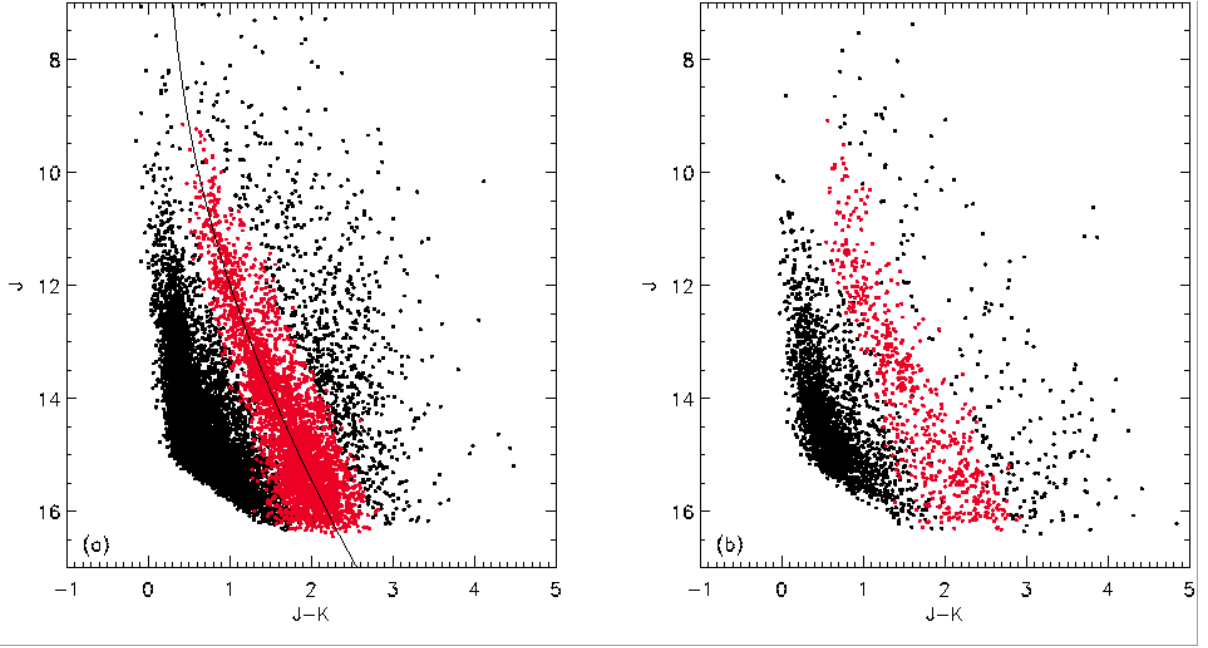}
\caption{\label{jjk284} The first panel (a) shows a color-magnitude
diagram for field sources (away from RCW~49) near \lsim~284\arcdeg.
Features common in Galactic sightlines are the vertically extended
concentrations of points at $J$-$K\simeq$0.5 corresponding to dwarf
stars and at 1$\lesssim J$-$K\lesssim$2 (red points) corresponding to
mostly red clump giants \citep[see text and][]{drimmel03}.  We can
isolate the red clump giants and fit their distribution (best fit is
the overplotted line) to determine \ajk{} and the absolute extinction
with distance ($J$ magnitudes per kpc, see text for discussion of
uncertainties).
The second panel (b) shows the plot for the region around the giant
\ion{H}{2} region RCW49 - interestingly, the spread of red clump
giants is very weak faintward of $J\simeq$14, $J$-$K\simeq$1.75.  This is
likely caused by an opaque interstellar cloud, and if that cloud is
associated with RCW49, we can estimate the distance to be
5$\pm$1~kpc.}
\end{figure}

In Figure \ref{jjk284}, the apparent J magnitudes of the field stars
near \lsim~284\arcdeg{} are plotted against their $J-K$ colors.  This
color-magnitude diagram shows features common along Galactic
sightlines, in particular those at longitudes where the disk component
dominates over the bulge component.  A vertically extended
concentration of points at $J-K \sim 0.5$ are predominantly
main-sequence dwarfs at a range of distances (8365 sources in this
particular plot).  A second concentration (3560 sources) extending
from $J\simeq 12, J-K\simeq 1$ to $J\simeq 16, J-K\simeq 2$ are
predominantly giants.  Distance spreads the distribution vertically in
the figure; extinction moves a statistically uniform sample of stars
vertically and horizontally by an amount proportional to distance,
provided that the dust per unit distance is approximately constant.
We note that the dwarf and giant sequences are not parallel -- this is
expected if the distribution of dwarfs is affected not only by
interstellar reddening at faint $J$ but also by the addition of
intrinsically faint, numerous cool dwarfs.  The absolute magnitudes of
giants, by contrast, are dominated by the red clump giants of early K
type, with fairly narrow color and luminosity distributions.  There
are no intrinsically faint local stars to be added to the giant
sequence at faint $J$; instead, we have a population of intrinsically
similar beacons, spread along the line of sight.  The distinct red
clump feature is discussed and exploited to probe Galactic structure
by \citet{lopez02}, and further used by \citet{drimmel03} to calculate
the amount of extinction along certain Galactic lines of sight.  The
859 points redward of the giants in Figure~\ref{jjk284} are probably
mostly intrinsically red asymptotic giant branch stars.

% FIGURE 3

We perform a 4-variable minimization to fit the curved red giant
branch star locus and determine the amount of extinction per unit
distance. The curve can be written parametrically as a function of
distance $d$:
\begin{eqnarray} 
  J &=& J_0 + 5\log\left({d/10\,\rm{pc}}\right) + c_J\left({d/10\,\rm{pc}}\right) \\
  J-K &=& J_0-K_0 + \left(c_J-c_K\right)\left({d/10\,\rm{pc}}\right), \label{jjkfit}
\end{eqnarray}
where $c_J$ is the average extinction per unit distance in the $J$
band, assumed constant for a given line of sight.  We select the red
giant clump stars (red points in Figure \ref{jjk284}) by their $JHK$
apparent magnitudes and colors, and use the IDL program {\tt amoeba}
to minimize the sum of the squared distances between each data point
and the line defined by Equation~\ref{jjkfit}.  
If we wish only to determine the
ratio $c_J/c_K = A_J/A_K$, then $J_0$ and $K_0$ can be determined by
best fit for the population of giants along the particular sightline.
We get the same results for $c_J/c_K$ whether $J_0$ and $K_0$ are
fixed at the values suggested by \citet{lopez02} for a K2III
($M_K=-1.65$, $H-K=0.75$, intrinsic spread of $0.3$ magnitudes in the
population), or allowed to vary.  We varied the region of the sky, the
color selection criteria for RGB stars, the initial guess for the four
parameters, and the degree of robustness in the fit algorithm
(iteratively rejecting high sigma outliers).  The range in fit
coefficients is larger than the numerically calculated sigma for each
fit, and we use that larger range as our quoted uncertainty.  The
uncertainty in the extinction per unit distance $c_J$ is larger than
in the ratio $c_J/c_K$, because it depends on assuming the absolute
$J$ and $K$ magnitudes of the red clump.

We fit the locus of red clump giants in all possible $JHK$
color-magnitude diagrams ($J-K$ versus $J$, $H-K$ versus $K$,
etc).  Our derived ratios are \ajk=2.5$\pm$0.2,
$A_H/A_K$=1.55$\pm$0.1, and $A_J/A_H$=1.65$\pm$0.1.  As noted above,
we also measure the average magnitudes of extinction per kpc along
this line of sight: $c_J$=0.35$\pm$0.15, $c_H$=0.25$\pm$0.1, and
$c_K$=0.15$\pm$0.1~mag~kpc$^{-1}$.  These are about 1$\sigma$ higher
than numbers often used for the average Galactic disk,
$c_V\simeq$0.7~mag~kpc$^{-1}$ or $c_J\simeq$0.2~mag~kpc$^{-1}$.
Examining the color-magnitude diagrams also reveals two regions of high
extinction along the line of sight towards \lsim~284\arcdeg: at
$J\simeq$12.7, $J-K\simeq$1.3 and $J\simeq$14.3, $J-K\simeq$1.7.
The extinction of the more distant cloud is so high towards the
RCW~49 giant \ion H2 region that few red clump giants are seen beyond
it (see Figure~\ref{jjk284}).  If that cloud is associated with RCW~49,
as seems reasonable, then we can estimate the distance to be
5$\pm$1~kpc, independently of any other distance measurement.
Finally, it is more difficult to fit the giant branch towards
\lsim~42\arcdeg{} because of greater contamination by AGB stars and the
bulge red giants at magnitudes fainter than $J \simeq 14$, but we find
the same ratios of $A_J/A_H/A_K$ within uncertainties.

%=========================================================================
\subsection{Determining color excess ratios}
\label{colex}

\begin{figure}
\epsscale{0.7}
\plotone{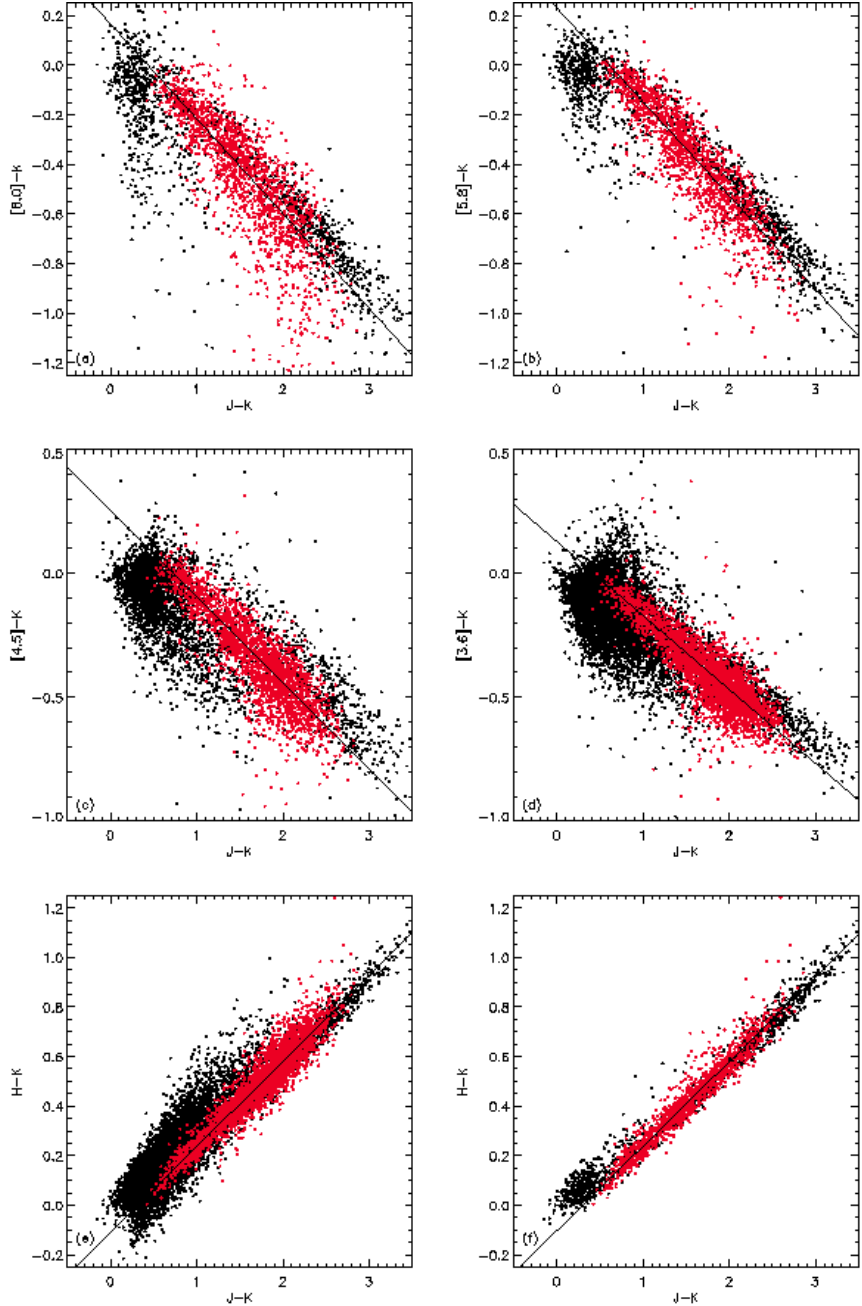}
\epsscale{1.0}
\caption{\label{2mccd}\small Color-color diagrams of sources detected
by \gl{} and 2MASS in the ``field'' ($b>0$, away from known \ion{H}{2}
regions) near \lsim~284\arcdeg.  The slope of a line fitted to each
source distribution is the color excess ratio \elk/\ejk.  We fit only
the red clump giants, selected by their location in the $J$, $J$-$K$
color-magnitude diagram and plotted here in red.  The first four
panels (a)-(d) show source colors $J$-$K$ versus [$\lambda$]-$K$ for
the four IRAC bands [$\lambda$].  Panels (d) and (e) show $J$-$K$
versus $H$-$K$, for sources with high-quality fluxes in the $JHK$
bands, and for only those sources with high-quality fluxes in all 7
filters, respectively.  If one fits only the latter stars, the slopes
change by $\lesssim$5\% and this has been included in error estimates.
The dominant source of scatter in these plots is variation in the
intrinsic colors of stars \citep[$\simeq\pm$0.3 mag][]{lopez02}.
Sources with nonphotospheric emission from circumstellar dust
(traditionally called ``infrared excess'' sources) appear as extreme
outliers, e.g. at the bottom of the first panel [8.0]-$K$ versus
$J-K$.}
\end{figure}

We measure each color excess ratio \elk/\ejk{} by fitting the locus of
a population of stars in a color-color diagram.  The color excess
\elk{} is defined as the difference between a star's observed and
intrinsic color. The excess for a particular star is
proportional to the amount of dust along the sightline to that star,
but ratios of color excesses reveal the wavelength dependence of extinction.
If one plots colors $[\lambda]-K$ versus $J-K$, the slope of the
source distribution is the color excess ratio \elk/\ejk, as
illustrated in Figure~\ref{2mccd}.

In each region of the sky, we fit a line to the stellar colors
$([\lambda]-K)/(J-K)$ of only the giant stars (selected by their $JHK$
magnitudes and plotted in red in Fig.~\ref{jjk284}).  Extremely red
sources (at the bottom of the first 4 panels in Figure~\ref{2mccd})
likely have nonphotospheric emission from circumstellar dust; these
outliers must be excluded from measurement of interstellar dust {\it
extinction}, because the dust {\it emission} has a different
wavelength dependence and appears as nonphysical extra reddening.  A
small number of highly evolved (AGB) stars can also fall in the
excluded region because of absorption bands in their atmospheres.
We calculated fits using several different algorithms, and used the
differences in results to estimate our uncertainties; the simplest fit
performed was a linear regression weighted by uncertainties in both
axes.  In addition to fitting all sources and fitting only red clump
giants, we tried arbitrarily excluding sources with extreme red excesses
(e.g. with $([\lambda]-K)<-0.3(J-K)-0.65$).  We also used a robust
algorithm which iterates the fit, rejecting high-sigma outliers.  This
also has the effect of excluding some of the extreme red excess
sources.

Figure~\ref{elam} shows the measured color excess ratios \elk/\ejk{}
plotted as a function of $\lambda^{-1}$ for the RCW~49 giant
\ion{H}{2} region, the field near \lsim~284\arcdeg{}, and the
field near \lsim~42\arcdeg.  The plotted values are listed in
Table~\ref{tab}, along with average values.  The uncertainties are a
combination of variation in fit results depending on the exact part of
the sky, stellar population, and fit algorithm, calculated uncertainty
of the fit coefficients, and errors in GLIMPSE and 2MASS photometry
(in order of decreasing significance).  There is good agreement among
the excess ratios measured in all three regions, although the data are
from very different directions in the Galactic plane.  More extensive
interpretation is possible after conversion of the measured color
excess ratios into relative extinction \alk{}, as discussed in the
next section.

% FIGURE 4

An alternative method for measuring the color excess ratios due to
interstellar extinction that is not dependent on the intrinsic stellar
color distribution is described by \citet{kenyon98}. The method
consists of comparing a selection of stars presumed to be mostly
behind an extincting cloud with a selection of stars not extinguished by
the cloud (``off-cloud'').  Instead of plotting colors (e.g., $H-K$),
one plots all possible color {\it differences} $([\lambda]-K)_i -
([\lambda]-K)_j$ versus $(H-K)_i - (H-K)_j$ (where $i$ indexes all
``cloud'' stars and $j$ all ``off-cloud'' stars).  If the stellar
color distributions are statistically the same in the two samples, the
color difference plot is a tight distribution extended along the
extinction vector.  As before, the slope of that distribution is the
color excess ratio \elk/\ejk. We calculated the wavelength dependence
of extinction from \gl{} and 2MASS data using the RCW~49 region as
``cloud'' and the region away from the \ion{H}{2} region to the north
as ``off-cloud.'' The results (\elk/\ejk{} = 0.36$\pm$0.02,
-0.31$\pm$0.02, -0.36$\pm$0.02, -0.40$\pm$0.03, and -0.37$\pm$0.02,
for $H$, [3.6], [4.5], [5.8], and [8.0], respectively) are the same
within the line-fitting uncertainties to the values listed in
Table~\ref{tab} that were obtained from the simpler method described
above.

We have discussed ratios of color excesses \elk/\ejk. The use of $J$
and $K$ in the denominator is convenient because we determine \ajk{}
from the color-magnitude relation (see \S\ref{jk}), but the \alk{} we
derive is independent of this choice.  For completeness, we fit other
color excess ratios, and found that the resulting ratios did not
change within uncertainty -- for example
\begin{eqnarray*} {{E_{[4.5]-K}}\over{E_{J-K}}} &=& 0.35\pm0.02, \\
\left({{E_{[4.5]-[3.6]}}\over{E_{[3.6]-K}}} + 1\right) \left({{E_{[3.6]-K}}\over{E_{J-K}}}\right) &=& \\
(0.16\pm0.03 + 1)(0.30\pm0.02) &=& 0.35\pm0.07
\end{eqnarray*}

%=====================================================================
\section{The wavelength dependence of interstellar extinction $A_{\lambda}$}
\label{discussion}

\subsection{Determination of \alk}

\begin{figure}
\plotone{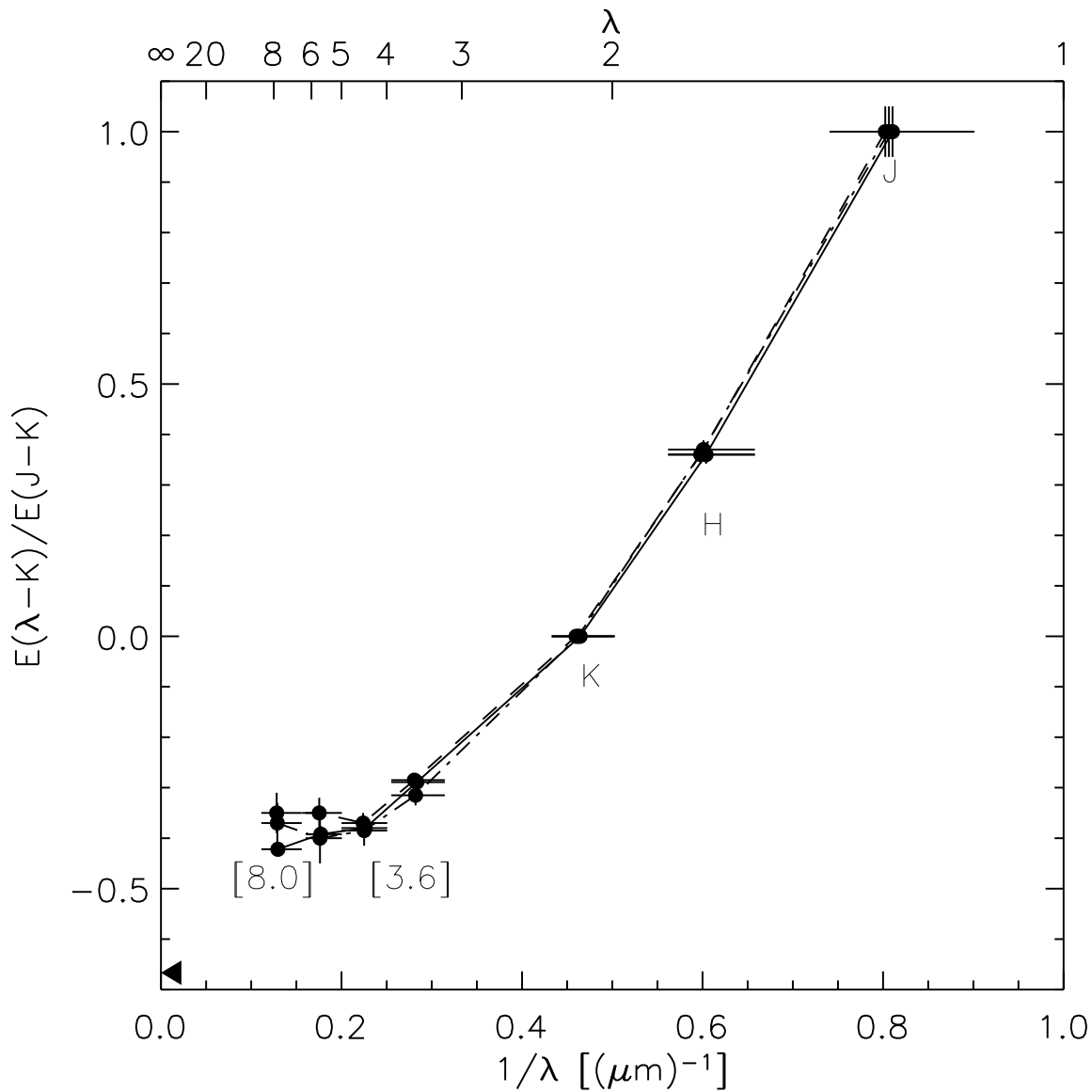}
\caption{\label{elam} The color excess ratio \elk/\ejk{} plotted as a
function of $1/\lambda$, for the RCW~49 region (dot-dashed) and for
the ``field'' near \lsim~284\arcdeg{} (solid), and for the ``field''
near \lsim~42\arcdeg{} (dashed). Error bars on the ``field'' curves
are smaller than for RCW~49, and comparable to the sizes of the
points.  Horizontal bars indicate the filter widths and the fact that
we have only sampled the wavelength dependence of interstellar
extinction at those discrete filters. The filled arrowhead is
discussed later in the text.}
\end{figure}

With color excess ratios from the slopes of color-color plots
(\S\ref{colex}) and \ajk{} determined from the color-magnitude
relation of giants (\S\ref{jk}), Equation \ref{alamk} yields \alk.
The presence of enough data points from 2MASS and GLIMPSE to measure
\ajk{} directly from our data is very important to this study.  Color
excess ratios are sensitive only to {\em changes} in the extinction
with wavelength.  Without \ajk, we would have to extrapolate
\elk/\ejk{} in Figure~\ref{elam} from $K$ band to $\lambda^{-1}=0$;
the intercept of that extrapolated curve is
$-A_K/(A_J-A_K)=1/(1-A_J/A_K)$.  Although this extrapolation is a
commonly used procedure, it is difficult to perform accurately, and
questionable because of the absorption features of silicates peaking
at 9.7 and 18$\mu$m.  The arrowhead at the lower left of
Figure~\ref{elam} shows the value of $1/(1-A_J/A_K)$ using our values
measured as described in section~\ref{jk}.  By assuming a power-law
through the $JHK$ relative extinctions (effectively an extrapolation
to $\lambda^{-1}=0$), \citet{mw90} estimated \ajk{} = 2.7. This value
requires the intercept in Figure~\ref{elam} to be -0.59.  Our
intercept derived from fitting the red clump giants is at -0.67.

\begin{figure}
\plottwo{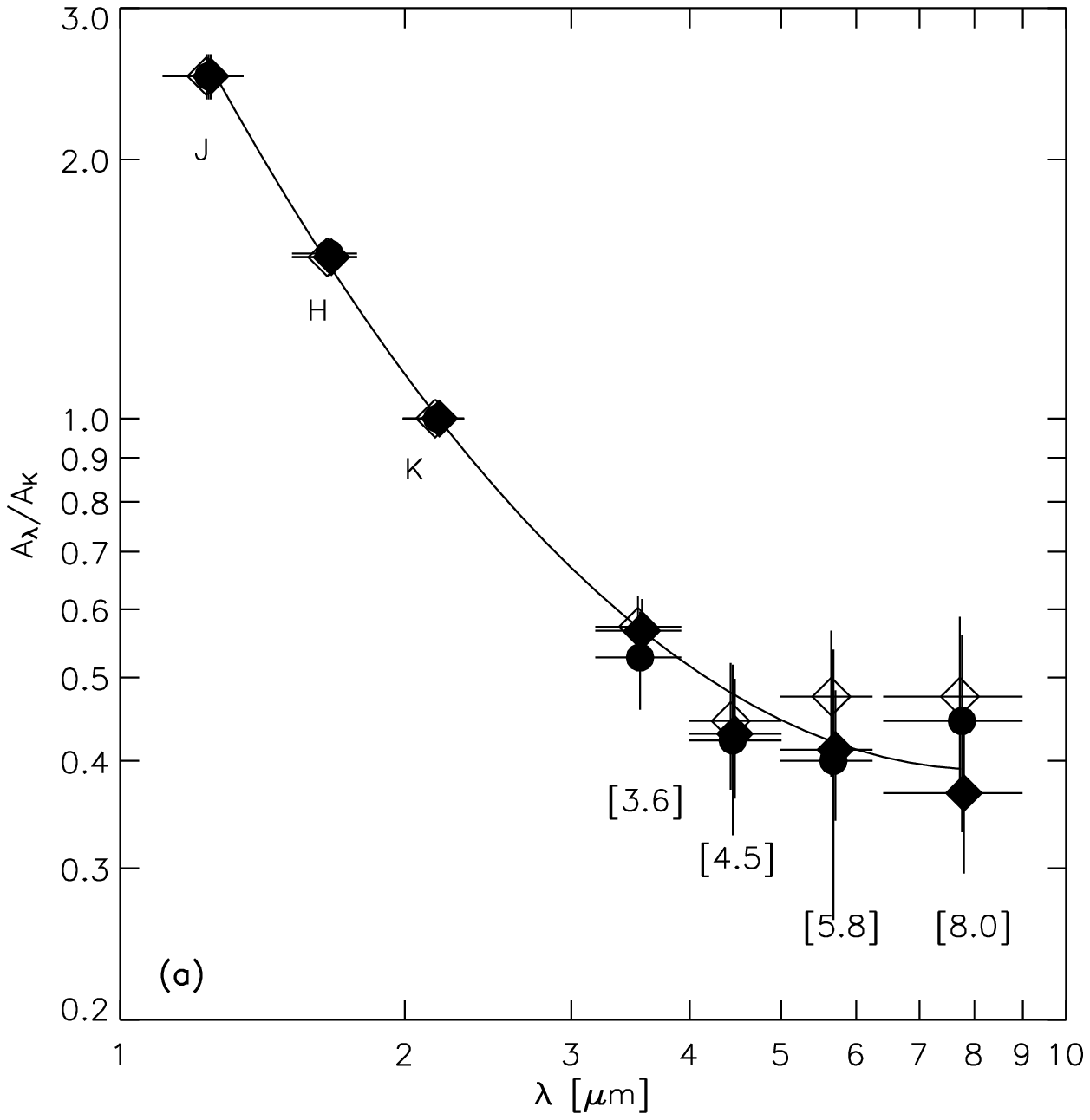}{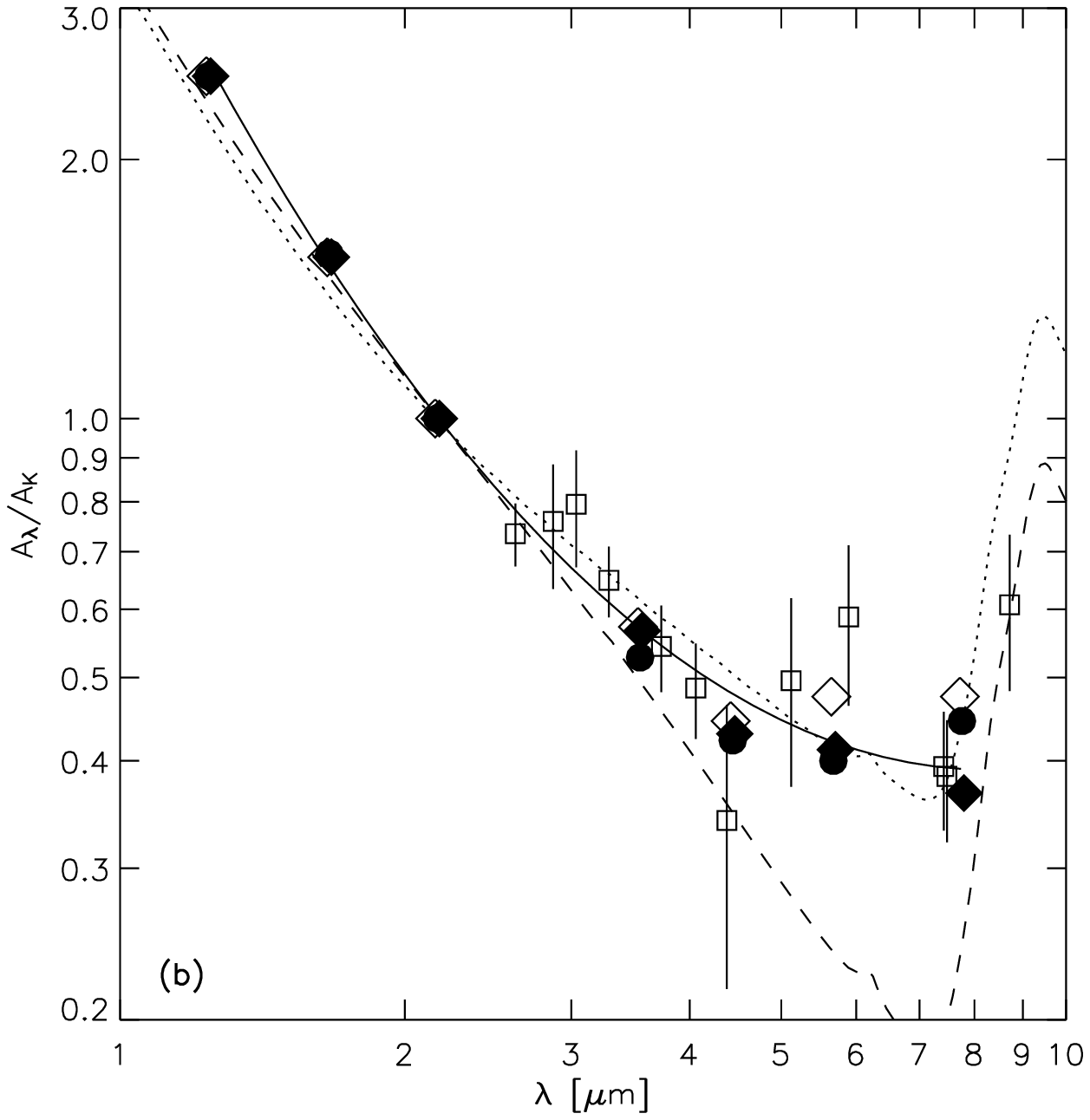}
\caption{\label{ak} In the first panel, symbols show the derived
relative extinction values for RCW~49 (solid circles), the field near
\lsim~284\arcdeg{} (solid diamonds), and the field near
\lsim~42\arcdeg{} (open diamonds).  The open and solid diamonds
coincide at $\lambda$=3.5$\mu$m.  The lines show fits (see
Equation~\ref{fit}) to the mean field (solid) and RCW~49 (dot-dashed).
Horizontal bars indicate the filter widths and the fact that we have
only sampled the wavelength dependence of interstellar extinction at
those discrete filters.  In the second panel, the measurements of
\citet{lutz} have been added (squares with error bars), as well as
theoretical curves from \citet{wd01} for $R_V$=5.5 (their case ``A''
dashed, their case ``B'' dotted).}
\end{figure}

Figure~\ref{ak} shows the derived \alk{} for the field stars at
\lsim~284\arcdeg{} (solid diamonds), for \lsim~42\arcdeg{} (open
diamonds), and for the dust surrounding RCW~49 (solid circles).  The
uncertainties are a combination of uncertainty in determining \ajk,
variation in fit results depending on the exact part of the sky,
stellar population, and fit algorithm, calculated uncertainty of the
fit coefficients, and errors in GLIMPSE and 2MASS photometry (in order
of decreasing significance).  There is good agreement between the
three determinations of \alk, in two very different directions of the
Galactic plane, and even towards a giant \ion{H}{2} region.

% FIGURE 6

A simple fit to the average \alk{} in the three regions is given by
\begin{equation}
\log\left[A_\lambda/A_K\right]=
0.61(\pm 0.04) - 2.22(\pm 0.17)\log(\lambda) + 1.21(\pm
0.23)(\log(\lambda))^2, \label{fit}
\end{equation}
where $\lambda$ is in \mic{}. The fit curve is plotted along with the
data points in Figure~\ref{ak} and is recommended for use between 1.25
and 7.75$\mu$m.  (Note that a simple power-law extinction curve with
$\beta$=1.8 has coefficients of 0.60, -1.8, and 0.0).

% TABLE
\begin{deluxetable}{l|ccccccc}
\tablecolumns{8}
\tablewidth{0pc}
\tablecaption{\label{tab} IRAC Extinction \alk}
\tablehead{
& \colhead{J} & \colhead{H} & \colhead{K} & 
\colhead{[3.6]} & \colhead{[4.5]} & \colhead{[5.8]} & \colhead{[8.0]} \\
& 1.240\mic\tablenotemark{a} & 1.664\mic & 2.164\mic & 3.545\mic & 4.442\mic & 5.675\mic & 7.760\mic
}
\startdata
& \multicolumn{7}{l}{\elk/\ejk} \\\tableline
RCW~49          & 1.0 & 0.37$\pm$0.02 & 0.0 & 0.31$\pm$0.03 & 0.38$\pm$.03 &
0.40$\pm$0.05 & 0.37$\pm$0.04 \\
\lsim~284\arcdeg& 1.0 & 0.35$\pm$0.02 & 0.0 & 0.30$\pm$0.02 & 0.35$\pm$.02 &
0.38$\pm$0.02 & 0.38$\pm$0.03 \\
\lsim~42\arcdeg & 1.0 & 0.36$\pm$0.02 & 0.0 & 0.29$\pm$0.02 & 0.37$\pm$.02 &
0.35$\pm$0.03 & 0.35$\pm$0.04 \\
average         & 1.0 & 0.36$\pm$0.02 & 0.0 & 0.30$\pm$0.01 & 0.37$\pm$.02 &
0.38$\pm$0.04 & 0.37$\pm$0.04 \\
\tableline & \multicolumn{7}{l}{\alk} \\\tableline
RCW~49          & 2.50$\pm$0.15 & 1.56$\pm$0.06 & 1.00 & 
0.53$\pm$0.07 & 0.42$\pm$0.09 & 0.40$\pm$0.14 & 0.45$\pm$0.11 \\
\lsim~284\arcdeg& 2.50$\pm$0.15 & 1.54$\pm$0.05 & 1.00 &
0.57$\pm$0.05 & 0.43$\pm$0.07 & 0.41$\pm$0.07 & 0.37$\pm$0.07 \\
\lsim~42\arcdeg & 2.50$\pm$0.15 & 1.54$\pm$0.06 & 1.00 & 
0.57$\pm$0.06 & 0.45$\pm$0.07 & 0.48$\pm$0.09 & 0.48$\pm$0.11 \\
average         & 2.50$\pm$0.15 & 1.55$\pm$0.08 & 1.00 &
0.56$\pm$0.06 & 0.43$\pm$0.08 & 0.43$\pm$0.10 & 0.43$\pm$0.10 \\ 
\enddata
\tablenotetext{a}{The adopted wavelengths are the isophotal
wavelengths of the 2MASS and IRAC filters convolved with a K2III
star - see \S\ref{intro}.}
\end{deluxetable}

For historical reasons, the filter to which \al{} is compared is
usually taken to be $A_V$, and our extinction measurements are
presented as \alk. In contrast to NIR extinction, optical extinction
($\lambda < 1$\mic) is known to vary significantly among sightlines
\citep{ccm89,fitz99}; the variation is well-characterized by a single
parameter, $R_V=A_V/E_{B-V}$. The variation of $A_V/A_K$ is beyond the
scope of this paper, but $A_V/A_K\sim$8.8 for $R_V$=3.1 and $\sim$7.5
for $R_V$=5 \citep{ccm89}. In general, the lack of variation of the
extinction curve throughout the NIR among various environments (in
comparison to its strong dependence on $R_V$ in the
optical/ultraviolet) suggests that it may not be possible to estimate
$A_V/A_K$ for our dataset from NIR observations alone.

%=====================================================================
\subsection{Comparison with previous measurements and discussion}

Previous measurements of the wavelength and spatial dependence of
interstellar extinction in this wavelength regime differ.
It is important to note that the measured excesses at IRAC wavelengths
($\lambda>3$\mic) lie above the curve extrapolated from
$\lambda<3$\mic (see Figure~\ref{elam}).  This implies a source of
extinction in the IRAC wavelength regime that is in excess of the
power-law wavelength dependence exhibited by a grain model tuned to
UV-optical extinction observed in the diffuse ISM \citep[e.g.][Case
A]{wd01}.
Our measurements reject a pure power-law extinction curve by
$>$4$\sigma$ at 6$\mu$m.
\citet{lutz} and \citet{lutz99} observed the Galactic Center with the Short Wavelength
Spectrometer on the {\it Infrared Space Observatory} (ISO) in the
interval 2.4--45\mic, using recombination lines of hydrogen to provide
intrinsic flux ratios. They converted their reddening values to
{$A_{\lambda}/A_K$} by assuming a theoretical value of $A_K/A_V$ from
\citet{dr89}.  Figure~\ref{ak} shows that the \citet{lutz} values are
very similar to those presented in this work but show more scatter.

Our values of \alk{} are consistent with \citet{lutz} and \citet{lutz99} in spite of the
large differences in method and sightlines.  Their color excess ratios
pertain to the line of sight to the Galactic center, while our
\lsim~42\arcdeg{} sightline probes only the outer 30\% of that
galactocentric distance and \lsim~284\arcdeg{} is confined to
approximately the solar circle.  The path to the Galactic center
contains $A_V\sim 5-10$ magnitudes of extinction from molecular clouds
\citep[which may exhibit different extinction than the diffuse ISM;
see discussion and references in][]{whittet97}, out of a total
$A_V\sim 25$ magnitudes.  The molecular fraction of the ISM in our
regions is at present unknown at adequate spatial resolution;
low-resolution $^{12}$CO \citep{dame} measurements suggest that only a
few magnitudes of the observed $A_V\sim 10-15$ may be from molecular
clouds.  Finally, using nebular emission lines to provide the
extinction assumes that the extinction is uniform over the
14$\arcsec\,\times\,20\arcsec$ aperture of the ISO spectrometer,
whereas using stellar photometry probes a pencil beam along each line
of sight.  Our measurements are consistent with \citet{lutz} despite
all of these differences in methodology and circumstance.

The similarities of the wavelength dependence of extinction for our
two very different Galactic longitudes and, to within larger errors,
towards the Galactic center \citep{lutz,lutz99}, is remarkable. This suggests
an almost universal extinction law in the infrared.  Even towards the
massive star-forming region RCW~49 the extinction is very similar to
that observed in the ``field'' regions.

It is useful to estimate the extent to which the [8.0] IRAC band is
affected by the silicate feature that peaks at 9.7\mic{}.  Most stars
emit the peak of their emission far shortward of the IRAC wavelengths,
and their emission in the Rayleigh-Jeans limit weights the filter
response towards shorter effective wavelengths.  The {\it Spitzer}
Science Center-provided isophotal wavelength is 7.735\mic{}, at which
the silicate opacity is $<$10\% of that at 9.7\mic{}
\citep[e.g.][]{jager94,jager98}. The ISO spectra of heavily obscured
objects in \citet{gibb04} show 8\mic{} fluxes that seem little
affected by silicate absorption. Nevertheless, the filter is broad,
and silicate absorption in the diffuse ISM appears consistent with a
FWHM of 2.3\mic, possibly wider in molecular clouds
\citep{roche,bowey}.  Convolution of the IRAC filter profile with a
source emitting a Rayleigh-Jeans spectrum and with a Gaussian
absorption feature at 9.7\mic{} shows that the silicate feature could
have as much as a 20\% effect on the filter flux.  Studies of
individual objects (especially those with circumstellar dust) could
show more variation in the [8.0] extinction with source spectral shape
than the average ISM results presented here.  The effect of the
silicate feature on the [5.8] IRAC band is negligible.

The ratio of polarization to extinction is an important diagnostic of the
nature of grains. Consider aligned grains of a particular type of material. At
a particular wavelength, both extinction and polarization are integrals of
similar optical constants over the size distribution of the grains. Both
polarizations and extinctions at both 4 and 8$\mu$m have been measured in two
molecular clouds: GL\,2591 and the BN object in Orion (see Martin \& Whittet
1990 for observations and references.) Because of atmospheric opacity, there
are no measurements between these wavelengths. In GL\,2591, $p$(4
$\mu$m)/$p$(8$\mu$m) $\sim$2.5; in BN, $p$(4$\mu$m)/$p$(8$\mu$m) $\sim$2.0.
The silicate strengths are also similar; $p$(10$\mu$m)/$p$(8$\mu$m) is
$\sim$3.7 for GL\,2591 and 4.3 for BN.  Equation (4) shows that our measured
extinction ratio $A(4 \mu{\rm m})/A(8 \mu{\rm m}) \sim$ 1.32, significantly
below the 4$\mu$m/8$\mu$m polarization ratio of either object. The
wavelength dependence of the polarization outside of the ice band is similar
for these two clouds but differs significantly from our measured wavelength
dependence of extinction, suggesting that {\em the grains providing the 8
$\mu$m extinction are different from the silicates at 10$\mu$m and those
providing 4$\mu$m extinction} (carbonaceous?), and that {\em they are not as
well aligned}. The alternative to this conclusion is that the dust within
Orion BN and GL\,2591 has a wavelength dependence of extinction significantly
different from the dust observed by GLIMPSE.

\section{Conclusions}
\label{conclusions}

We have measured the wavelength dependence of interstellar extinction
in the 1.25--8.0\mic{} region by combining \gl{} and 2MASS
observations.  The values calculated for two very different directions
in the Galactic plane are remarkably similar.  Even near the giant
\ion{H}{2} region RCW~49, extinction is not significantly different.
It may be possible to study variations of extinction with Galactic
longitude using the entire \gl{} survey, expected to be complete by
early in the year 2005.  Since the values of \ejhk{} in the Galactic
plane are somewhat higher than at higher Galactic latitude, there may
be differences in the extinction at IRAC wavelengths as well.

The similar derived behavior of the extinction along several different
sightlines suggests that our relative extinction values \alk{} may be
generally applicable.  Our average extinction measurements may be used
to correct \gl{} and other IRAC data in the Galactic plane.  These
values are provided in Table~\ref{tab}, and the simple formula of
Equation~\ref{fit} provides an analytical fit of simple functional
form.

\acknowledgments

We acknowledge the invaluable assistance of Stephan Jansen for
maintaining the GLIMPSE computing network.  We thank the referee,
Dr. Bruce Draine, for comments and discussion which improved the paper.
Support for this work, part of the {\it Spitzer Space Telescope}
Legacy Science Program, was provided by NASA through Contract Numbers
(institutions) 1224653 (UW), 1225025 (BU), 1224681 (UMd), 1224988
(SSI), 1242593 (UCB), 1253153 (UMn), 11253604 (UWy), 1256801 (UWW) by
the Jet Propulsion Laboratory, California Institute of Technology
under NASA contract 1407.
This research made use of {\sc MONTAGE}, funded by NASA's Earth
Science Technology Office, Computational Technologies Project, under
Cooperative Agreement Number NCC5-626 between NASA and the California
Institute of Technology. We acknowledge use of data products from the
Two Micron All Sky Survey, a joint project of the University of
Massachusetts and the Infrared Processing and Analysis
Center/California Institute of Technology, funded by NASA and the
NSF. {\it Spitzer} data used in this paper are from PID~195 (multiple
AORs) and from PID 631, AOR~0007283968

%=============================================================================

\end{document}